\documentclass[11pt]{amsart}

\usepackage{amsthm}

  \newtheorem{theorem}{Theorem}

  \newtheorem{remark}{Remark}
  \newtheorem{example}{Example}

  \newcommand {\pf}  {\mbox{\sc Proof. \,\,}}

\begin {document}

\title[Continued fractions and RSA]{Continued fractions and RSA with small secret exponent }

\author
{{\sc Andrej Dujella}  }

\date{}

\begin{abstract}
\noindent Extending the classical Legendre's result, we describe
all solutions of the inequality $|\alpha - a/b| < c/b^2$ in terms
of convergents of continued fraction expansion of $\alpha$.
Namely, we show that $a/b = (rp_{m+1} \pm sp_m) / (rq_{m+1} \pm
sq_m)$ for some nonnegative integers $m,r,s$ such that $rs < 2c$.
As an application of this result, we describe a modification of
Verheul and van Tilborg variant of Wiener's attack on RSA
cryptosystem with small secret exponent.
\end{abstract}

\maketitle


\section{Introduction}

The most popular public key cryptosystem in use today is
the RSA \cite{RSA}. Its security is based on
the difficulty of finding the prime factors of large
integers.

The modulus $n$ of a RSA cryptosystem is the product of two large
primes $p$ and $q$. The public exponent $e$ and the secret
exponent $d$ are related by $ed\equiv 1 \pmod{\varphi(n)}$, where
$\varphi(n)= (p-1)(q-1)=n-p-q+1$. In a typical RSA cryptosystem
$p$ and $q$ have approximately the same number of bits, and $e<n$.
The encryption and decryption algorithms are given by $C= M^e
\bmod n$, $M=C^d \bmod n$.

To speed up the RSA encryption or decryption one may try to use small
public or secret decryption exponent.
The choice of a small $e$ or $d$ is especially
interesting when there is a large difference in computing power
between two communicating devices,
e.g. in communication between a smart card and a larger computer.
In this situation,
it would be desirable for the
smart card to have a small secret exponent, and for the
larger computer to have a small public exponent in order
to reduce the processing required in the smart card.

However, in 1990 Wiener \cite{Wiener} described an attack on a
typical RSA with small secret exponent. He showed that if
$d<n^{0.25}$, then $d$ is the denominator of some convergent of
the continued fraction expansion of $e/n$, and therefore $d$ can
be computed efficiently from the public key $(n,e)$. His result is
based on the classical Legendre's theorem on Diophantine
approximations of the form $|\alpha - \frac{a}{b}| <
\frac{1}{2b^2}$. Pinch \cite{Pinch} extended the attack to some
other cryptosystems. In 1997, Verheul and van Tilborg proposed an
extension of Wiener's attack that allows the RSA cryptosystem to
be broken by an exhaustive search when $d$ is a few bits longer
than $n^{0.25}$.

In this paper, we will generalize Legendre's result to Diophantine
approximations of the form $|\alpha - \frac{a}{b}| <
\frac{c}{b^2}$. We will show that this result leads to the more
efficient variant of the above mentioned attacks.

\bigskip

Our attack on RSA will closely follow Wiener's ideas, but let us
very briefly mention some other attacks on RSA with small exponent
$d$. In 1999, Boneh and Durfee \cite{B-D1} proposed an attack on
RSA with small secret exponent which is based on Coppersmith's
lattice-based technique for finding small roots of bivariate
modular polynomial equation \cite{Cop}. The attack works if $d<
n^{0.292}$. Similar attack was proposed Bl\"omer and May if $d<
n^{0.29}$. Recently, it was noted by Hinek, Low and Teske
\cite{H-L-T} (see also \cite{Hinek}) that these theoretical bounds
on $d$ are not correct (some quantity which appears in the
analysis is not negligible). Also, it should be noted that the
Coppersmith's theorem is for univariate case; in the bivariate
case it is only a heuristic result for now. On the other hand, it
seems that these attacks work well in practice.

\section{Wiener's attack on RSA} \label{sec:wie}

In 1990, Wiener \cite{Wiener} described a polynomial time
algorithm for breaking a typical (i.e. $p$ and $q$ are of the
same size and
$e<n$) RSA cryptosystem if the secret exponent $d$ has at most
one-quarter as many bits as the modulus $n$. The Wiener's attack
is usually described in the following form (see
\cite{B-notices,Smart}):

If $p<q<2p$, $e<n$ and $d<\frac{1}{3}\sqrt[4]{n}$, then $d$ is the
denominator of a convergent of the continued fraction expansion of
$\frac{e}{n}$.

The starting point is the basic relation between exponents
$$ed\equiv 1 \pmod{\varphi(n)}.$$ This means that there is an
integer $k$ such that $ed- k\varphi(n)=1$. Now, $ \varphi(n)
\approx n$ implies $\frac{k}{d} \approx \frac{e}{n}$. More
precisely, we have $ n-3\sqrt{n} < \varphi(n) < n$ and $$ \Big|
\frac{k}{d} - \frac{e}{n} \Big| < \frac{3k}{d\sqrt{n}} <
\frac{1}{2d^2}. $$ Hence, by Legendre's theorem, $\frac{k}{d}$ is
a convergent of continued fraction expansion of $\frac{e}{n}$.

If $[a_0;a_1,a_2, ...]$ is the continued fraction expansion
of a real number $\alpha$, then the convergents $\frac{p_j}{q_j}$
satisfy $p_0=a_0$, $q_0=1$, $p_1=a_0a_1+1$, $q_1=a_1$,
\begin{eqnarray*}
p_i &=& a_ip_{i-1}+p_{i-2}, \\ q_i &=& a_iq_{i-1}+q_{i-2}.
\end{eqnarray*}
Therefore, the denominators grow exponentially. This means that
total number of convergents of $\frac{e}{n}$ is of order $O(\log{n})$.
If a convergent can be tested in polynomial time, this
will give us a polynomial algorithm to determine $d$.

Wiener proposed the following method for testing convergents. Let
$\frac{a}{b}$ be a convergent of $\frac{e}{n}$. If it is the
correct guess for $\frac{k}{d}$, than $\varphi(n)$ can be computed
from $\varphi(n) = (p-1)(q-1)= (be-1)/a$. Now we can compute
$\frac{p+q}{2}$ from the identity $$ \frac{pq - (p-1)(q-1) +1}{2}
= \frac{p+q}{2}, $$ and $\frac{q-p}{2}$ from the identity $(
\frac{p+q}{2} )^2 - pq = ( \frac{q-p}{2} )^2$. If the numbers
$\frac{p+q}{2}$ and $\frac{q-p}{2}$, obtained by these identities,
are positive integers, then the convergent $\frac{a}{b}$ is
correct guess for $\frac{k}{d}$. We can also recover easily $p$
and $q$ from $\frac{p+q}{2}$ and $\frac{q-p}{2}$.

Another possibility for detecting the correct convergent is by
testing which one gives a $d$ which satisfies $(M^e)^d =
M\pmod{n}$ for some random value of $M$.

\begin{example} \label{ex:1}
{\rm Let $n=7978886869909$, $e=3594320245477$, and assume that
$d<561$. Continued fraction expansion of $\frac{e}{n}$ is $$  [0;
2, 4, 1, 1, 4, 1, 2, 31, 21, 1, 3, 1, 16, 3, 1, 114, 10, 1, 4, 5,
1, 2], $$ and the convergents are $$ 0,\, \frac{1}{2},\,
\frac{4}{9},\,  \frac{5}{11},\, \frac{9}{20},\, \frac{41}{91},\,
\frac{50}{111},\, \frac{141}{313},\, \frac{4421}{9814},\, \ldots
\,.$$ Applying test $(2^e)^d \equiv 2 \pmod{n}$, we obtain
$d=313$. Of course, the same result can be obtained with the
original Wiener's test. For $\frac{a}{b}=\frac{141}{313}$ we find
$\frac{p+q}{2}= 2878805$, $\frac{q-p}{2}=555546$, and this yields
the factorization $n=2323259 \cdot 3434351$ }
\end{example}

\bigskip

We have seen in the previous example that the correct convergent
was the last convergent with denominator less than
$\frac{1}{3}\sqrt[4]{n}$. This suggests that perhaps it is not
necessary to test all convergents. We will justify this assertion.

To do that, we need more precise estimate of $| \frac{k}{d} -
\frac{e}{n}|$, which corresponds to better approximation of
$\varphi(n)$. Assume that $p < q < 2p$. Then $\frac{(p+q)^2}{n} =
2+ \frac{p^2+q^2}{pq}$ and thus $2\sqrt{n} < p+q <
\frac{3\sqrt{2}}{2} \sqrt{n} < 2.1214 \sqrt{n}$. This implies $$
\frac{k}{d} - \frac{e}{n} = \frac{k(p+q)-k-1}{dn} >
\frac{2k(\sqrt{n}-1)}{dn}. $$ Since $\frac{k}{d} > \frac{e}{n}
\cdot \frac{n}{n-2\sqrt{n}+1}$, we obtain
\begin{equation} \label{>}
  \frac{k}{d} - \frac{e}{n}  > \frac{2e}{n\sqrt{n}}.
\end{equation}
In the opposite direction we have
$$  \frac{k}{d} - \frac{e}{n} < \frac{2.1214 k}{d\sqrt{n}}. $$
We may assume that $n>{10}^8$. Then $\frac{k}{d} < 1.00023 \frac{e}{n}$,
and finally
\begin{equation} \label{<}
  \frac{k}{d} - \frac{e}{n} < \frac{2.122\, e}{n\sqrt{n}}.
\end{equation}
Similarly we find that
$$ \frac{k}{d} - \frac{e}{n} <
\frac{3.183\, e}{n\sqrt{n}}$$
if $p < q < 8p$.

In the rest of the paper we will work under the assumption that
$p<q<2p$, but the arguments can be easily modified to the case
$p<q<8p$.

From (\ref{>}) and (\ref{<})  we may conclude that $\frac{k}{d}$
is unique (odd) convergent satisfying $$ \frac{2e}{n\sqrt{n}} <
\frac{k}{d} - \frac{e}{n} < \frac{2.122\, e}{n\sqrt{n}}. $$
Indeed, this follows from the fact that if $p_m/q_m$ and
$p_{m+2}/q_{m+2}$ are two successive (odd) convergents of a real
number $\alpha$, then $p_{m+2}/q_{m+2}$ at least twice better
approximation of $\alpha$ than $p_{m}/q_{m}$, which is direct
consequence of the following well-known property of convergents
(see \cite[Theorems 9 and 13]{Hin})
\begin{equation} \label{hin}
\frac{1}{q_m(q_{m+1}+q_m)} <
\Big| \alpha -\frac{p_m}{q_m} \Big| < \frac{1}{q_m q_{m+1}} .
\end{equation}
Furthermore, if $\frac{k}{d}=\frac{p_m}{q_m}$, then $$
\frac{n\sqrt{n}}{4.244e} < q_mq_{m+1} < \frac{n\sqrt{n}}{2e}, $$
and $m$ is the unique odd positive integer satisfying this
inequality. This observations lead to an efficient algorithm for
finding the correct convergent in the Wiener's attack. Namely,
$\frac{k}{d}=\frac{p_m}{q_m}$, where $m$ is the smallest odd
positive integer such that $q_mq_{m+1} >
\frac{n\sqrt{n}}{4.244e}.$

\bigskip

As suggested in the original Wiener's paper, the attack can be
slightly improved by using better approximation to $\frac{k}{d}$,
e.g. $\frac{e}{f}$, where $f=n-\lfloor 2\sqrt{n} \rfloor +1$. This
can be combined with known extensions of Legendre's theorem.
Namely, there is an old result of Fatou \cite{Fatou} (see also
\cite[p. 16]{Lang}) which says that if $| \alpha - \frac{a}{b}| <
\frac{1}{b^2}$, then $\frac{a}{b} = \frac{p_{m}}{q_{m}}$ or
$\frac{p_{m+1}\pm p_{m}}{q_{m+1}\pm q_{m}}$. In 1981, Worley
\cite{Wor} (see also \cite{D-J} and \cite{O-L-W}) proved that $|
\alpha - \frac{a}{b}| < \frac{2}{b^2}$ implies $\frac{a}{b} =
\frac{p_{m}}{q_{m}}$, $\frac{p_{m+1}\pm p_{m}}{q_{m+1}\pm q_{m}}$,
$\frac{2p_{m+1}\pm p_{m}}{2q_{m+1}\pm q_{m}}$, $\frac{3p_{m+1}+
p_{m}}{3q_{m+1}+ q_{m}}$, $\frac{p_{m+1}\pm 2p_{m}}{q_{m+1}\pm
2q_{m}}$ or $\frac{p_{m+1}- 3p_{m}}{q_{m+1}- 3q_{m}}$.

We have $$ 0< \frac{k}{d} - \frac{e}{f} <
\frac{0.1221}{\sqrt{n}}.$$ If $d<4.04 \sqrt[4]{n}$, then
$\frac{0.1221}{\sqrt{n}} < \frac{2}{d^2}$ and $d$ can be found in
polynomial time (which extends the Wiener's attack by the factor
12).

More general extensions of Wiener's attack will be considered in
next sections.

\section{Verheul and van Tilborg variant of Wiener's attack} \label{sec:VT}

In 1997,  Verheul and van Tilborg \cite{V-vT} proposed the
following extension of Wiener's attack.

Let $m$ be the largest (odd) integer satisfying
$\frac{p_m}{q_m} - \frac{e}{n} > \frac{2.122\, e}{n\sqrt{n}}$.
Search for $\frac{k}{d}$ between fractions of the form
$\frac{r p_{m+1}+s p_m}{rq_{m+1}+ s q_m}$ ,
i.e. consider the system
\begin{eqnarray*}
r p_{m+1}+s p_m &=& k \\
r q_{m+1}+ s q_m &=& d.
\end{eqnarray*}
The determinant of the system satisfies
$|p_{m+1}q_m - q_{m+1}p_m| = 1$, and therefore
the system has (positive) integer solutions:
\begin{eqnarray*}
r &=& dp_{m} -k q_m  \\
s &=& kq_{m+1}- d p_{m+1} .
\end{eqnarray*}
If $r$ and $s$ are small, then they can be found
by an exhaustive search.

\bigskip

Let us estimate the number of steps in this exhaustive search,
i.e. let us find upper bounds for $r$ and $s$. Let
$d=D\sqrt[4]{n}$.

From (\ref{hin}) it follows $r = dq_m \Big(\frac{p_m}{q_m} -
\frac{k}{d} \Big) < \frac{d}{q_{m+1}}$. The estimate for $s$
depends on the sign of the number $\frac{e}{n} -
\frac{p_{m+1}}{q_{m+1}} - \frac{2.122e}{n\sqrt{n}}$. (We may
expect that this number will be positive in 50\% of the cases.)
Assume that $\frac{e}{n} - \frac{p_{m+1}}{q_{m+1}} >
\frac{2.122e}{n\sqrt{n}}$. Then $$s = dq_{m+1} \Big(\frac{k}{d} -
\frac{p_{m+1}}{q_{m+1}} \Big) < 2dq_{m+1} \Big(\frac{e}{n} -
\frac{p_{m+1}}{q_{m+1}} \Big) < \frac{2d}{q_{m+2}}. $$ Since $$
\frac{1}{q_{m+2}^2(a_{m+3}+2)} < \frac{p_{m+2}}{q_{m+2}} -
\frac{e}{n} < \frac{2.122 e}{n\sqrt{n}} <
\frac{2.122}{\sqrt{n}},$$ we have $$q_{m+2} >
\frac{\sqrt[4]{n}}{\sqrt{2.122(a_{m+3}+2)}}.$$ Also,  $q_{m+1} >
\frac{q_{m+2}}{a_{m+2}+1}$. Putting all these estimates together
we obtain
\begin{eqnarray*}
r &<& \sqrt{2.122(a_{m+3}+2)}(a_{m+2}+1)D, \\
s &<& \sqrt{2.122(a_{m+3}+2)}D.
\end{eqnarray*}
Hence, in this case the number of steps is bounded by
$$2.122(a_{m+3}+2)(a_{m+2}+1)D^2.$$

\medskip

Assume now that $\frac{e}{n} - \frac{p_{m+1}}{q_{m+1}} \leq
\frac{2.122e}{n\sqrt{n}}$. Then $$s = dq_{m+1} \Big(\frac{k}{d} -
\frac{p_{m+1}}{q_{m+1}} \Big) < dq_{m+1} \Big(\frac{p_m}{q_m} -
\frac{p_{m+1}}{q_{m+1}} \Big) = \frac{d}{q_{m}}.$$ Since in this
case is already $\frac{p_{m+1}}{q_{m+1}}$ close enough to
$\frac{e}{n}$, we have the estimate for $q_{m+1}$ which is
analogous to the estimate for $q_{m+2}$ in the previous case: $$
q_{m+1} > \frac{\sqrt[4]{n}}{\sqrt{2.122(a_{m+2}+2)}}.$$
This implies
\begin{eqnarray*}
r &<& \sqrt{2.122(a_{m+2}+2)}D, \\
s &<& \sqrt{2.122(a_{m+2}+2)}(a_{m+1}+1)D
\end{eqnarray*}
and in this case the number of steps is bounded by
$$2.122(a_{m+2}+2)(a_{m+1}+1)D^2.$$

\bigskip

In \cite{V-vT}, the authors propose that with reasonable
probability (20\%) the number of steps can be bounded by $256D^2$.
It is indeed true if we have in mind that partial quotients
$a_i$'s are usually very small. In \cite[p. 352]{Knuth} the
distribution of the partial quotients of a random real number
$\alpha$ is given. Approximately, $a_i$ will be 1 with probability
41.5\%, $a_i=2$ with probability 17.0\%, $a_i=3$ with probability
9.3\%, $a_i=4$ with probability 5.9\%, etc. Our analysis shows
that the success of Verheul and van Tilborg attack (when $D^2$ is
of reasonable size) depends heavily on the size of corresponding
partial quotients $a_{m+1}$, $a_{m+2}$ and $a_{m+3}$. And although
they are usually small, we cannot exclude the possibility that at
least one of them is large (see Examples \ref{ex:2} and
\ref{ex:3}). Namely, the probability that $a_i \geq x$ is equal to
$ \log_2(1+\frac{1}{x}),$ and this is a slowly decreasing
function.

In Section \ref{sec:mod} we will propose a method how to overcome
this problem and remove the dependence on partial quotients. A
general result on Diophantine approximation from the next section
will allow us to obtain more precise information on $r$ and $s$
which will reduce the number of steps in the search.

\section{Extension of Legendre's theorem}

\begin{theorem} \label{tm:kbb}
Let $\alpha$ be an irrational number and let $a$, $b$ be coprime
nonzero integers, satisfying the inequality
\begin{equation} \label{kb2}
 \Big| \alpha - \frac{a}{b} \Big| < \frac{c}{b^2},
\end{equation}
where $c$ is a positive real number. Then $(a,b) = (rp_{m+1} \pm
sp_{m},rq_{m+1} \pm sq_{m})$, for some nonnegative integers $m$,
$r$ and $s$ such that $rs < 2c$.
\end{theorem}

\pf Assume that $\alpha < \frac{a}{b}$, the other case is
completely analogous. Let $m$ be the largest odd integer
satisfying $$ \alpha < \frac{a}{b} \leq \frac{p_m}{q_m}. $$ If
$\frac{a}{b} > \frac{p_1}{q_1}$, we will take $m=-1$, following
the convention that $p_{-1}=1$, $q_{-1}=0$.

Let us define the numbers $r$ and $s$ by:
\begin{eqnarray*}
a &=& rp_{m+1} + sp_{m}, \\
b &=& rq_{m+1} + sq_{m}.
\end{eqnarray*}
Since $|p_{m+1}q_m - p_mq_{m+1}|=1$, we conclude that $r$ and $s$
are integers, and since $\frac{p_{m+1}}{q_{m+1}} < \frac{a}{b}
\leq \frac{p_m}{q_m}$, we have that $r\geq 0$ and $s>0$.

From the maximality of $m$, we have that $$ \Big|
\frac{p_{m+2}}{q_{m+2}} - \frac{a}{b} \Big| < \Big| \alpha -
\frac{a}{b} \Big| < \frac{c}{b^2}. $$ But {\small
\begin{eqnarray*}
\Big| \frac{p_{m+2}}{q_{m+2}} - \frac{a}{b} \Big| &\!=\!&
\frac{(a_{m+2}q_{m+1}\!+\!q_m)(rp_{m+1}\!+\!sp_m) -
(a_{m+2}p_{m+1}\!+\!p_m)(rq_{m+1}\!+\!sq_m)}{bq_{m+2}} \\ &\!=\!&
\frac{sa_{m+2}-r}{bq_{m+2}}.
\end{eqnarray*}}
Therefore, we obtain $$ b(sa_{m+2}-r) < cq_{m+2}=
\frac{c}{s}((sa_{m+2}-r)q_{m+1} + b), $$ which implies
$$(sa_{m+2}-r)(b-\frac{c}{s} q_{m+1}) < \frac{c}{s} \, b. $$
Furthermore we have $$ \frac{1}{sa_{m+2}-r} >
\frac{b-\frac{c}{s}q_{m+1}}{\frac{c}{s}b} = \frac{s}{c}-
\frac{1}{r+\frac{sq_m}{q_{m+1}}} \geq \frac{s}{c}- \frac{1}{r}. $$
Therefore, we obtain the following inequality
\begin{equation} \label{rsin}
r^2 -sr a_{m+2} +c a_{m+2} >0.
\end{equation}
We will consider (\ref{rsin}) as a quadratic inequality in $r$.

Assume for a moment that $s^2a_{m+2}\geq 4c$. Then $s^4a_{m+2}^2 -
4cs^2a_{m+2} \geq (s^2a_{m+2} -4c)^2$, and therefore (\ref{rsin})
implies
\[ r <\frac{1}{2s}\Big( s^2 a_{m+2}-\sqrt{s^4 a_{m+2}^2-
4cs^2a_{m+2}}\Big) \leq \frac{2c}{s}, \] or
\[ r >\frac{1}{2s}\Big( s^2a_{m+2}+\sqrt{s^4a_{m+2}^2- 4cs^2a_{m+2}}\Big)
\geq
\frac{1}{s}\Big(s^2a_{m+2} -2c). \]
The first possibility gives us the condition $rs < 2c$,
as claimed in the theorem.

Let us consider the second possibility, i.e.
\begin{equation} \label{second}
rs > s^2a_{m+2} -2c.
\end{equation}
Let us define $t=sa_{m+2} - r$. Since $\frac{p_{m+2}}{q_{m+2}} <
\frac{a}{b}$, we conclude that $t$ is a positive integer. Now we
have
\begin{eqnarray*}
a &=& rp_{m+1} + sp_{m} = (sa_{m+2}-t)p_{m+1} + sp_m = sp_{m+2}-
tp_{m+1}, \\ b &=& rq_{m+1} + sq_{m} = (sa_{m+2}-t)q_{m+1} + sq_m
= sq_{m+2}- tq_{m+1},
\end{eqnarray*}
and the condition (\ref{second}) becomes $st < 2c$.

Hence we proved the statement of the theorem under assumption
that $s^2a_{m+2}\geq 4c$.

Assume now that $s^2a_{m+2} < 4c$. Since $r<sa_{m+2}$, we have
two possibilities. If $r< \frac{1}{2} sa_{m+2}$, then
$rs < \frac{1}{2} s^2a_{n+2} < 2c$, and if
$r\geq \frac{1}{2} sa_{m+2}$, then $t=sa_{m+2}-r \leq
\frac{1}{2} sa_{m+2}$ and $st \leq \frac{1}{2} s^2a_{m+2} < 2c$.

\qed

\begin{remark}
{\rm It is not clear from the proof whether above theorem is valid
for rationals $\frac{a}{b}$ such that $\frac{a}{b} <
\frac{p_0}{q_0}= \lfloor \alpha \rfloor$. But this case
corresponds to the minus case with $m=0$ is the statement of the
theorem. Indeed, let $\frac{s}{r} = \lfloor \alpha \rfloor -
\frac{a}{b}$. Then $\frac{a}{b} = p_0 - \frac{s}{r} = \frac{rp_0 -
s}{r} = \frac{rp_0 - sp_{-1}}{rq_0 - sq_{-1}}$, and $rs = b^2\cdot
\frac{s}{r} < b^2\cdot \frac{c}{b^2} = c$. }
\end{remark}

\begin{remark}
{\rm The statement of the theorem is valid also for rational
numbers $\alpha$. Indeed, if $\alpha \in \mathbb{Q}$, then there
exist an integer $j\geq 0$ such that $\alpha = \frac{p_j}{q_j}$.
The proof is identical as in the irrational case, unless $\alpha <
\frac{a}{b} < \frac{p_{j-1}}{q_{j-1}}$ (or $\alpha > \frac{a}{b} >
\frac{p_{j-1}}{q_{j-1}}$). If we define positive integers $r$ and
$s$ by
\begin{eqnarray*}
a &=& rp_{j} + sp_{j-1}, \\
b &=& rq_{j} + sq_{j-1},
\end{eqnarray*}
then the inequalities $\Big| \alpha - \frac{a}{b}\Big| =
\frac{s}{bq_j} < \frac{c}{b^2}$ and $b>rq_j$ imply $rsq_j < sb <
cq_j$, and finally $rs < c$. }
\end{remark}

\bigskip

Similar result as our Theorem \ref{tm:kbb} was proved, with
different methods, by Worley. In \cite[Theorem 1]{Wor}, it was
shown that there are three types of solutions of the inequality
(\ref{kb2}). Two types correspond to $+$ and $-$ signs in
$(rp_{m+1} \pm sp_{m},rq_{m+1} \pm sq_{m})$, while Theorem
\ref{tm:kbb} shows that the third type can be omitted.

\medskip

Theorem \ref{tm:kbb} extends results for $c=1$ and $c=2$ cited in
Section \ref{sec:wie}. The result for $c=2$ has already found
applications in solving some Diophantine equations. In
\cite{O-L-W}, it is applied to the problem of finding positive
integers $a$ and $b$ such that $(a^2+b^2)/(ab+1)$ is an integer,
and in \cite{D-J} it is used for solving the family of Thue
inequalities $$|x^4 - 4cx^3y+(6c+2)x^2y^2 + 4cxy^2+y^4| \leq 6c+4.
$$ We hope that Theorem \ref{tm:kbb} will also find its
application in Diophantine analysis.

\section{A variant of Wiener's attack} \label{sec:mod}

In this section we propose new variant of Wiener's attack. It is
very similar to Verheul and van Tilborg attack, but instead of
exhaustive search after finding the appropriate starting
convergent, this new variant also uses estimates which follow from
Diophantine approximation (Theorem \ref{tm:kbb}).

\medskip

Let $m$ be the largest (odd) integer such that $$ \frac{p_m}{q_m}
> \frac{e}{n} + \frac{2.122 e}{n\sqrt{n}}. $$ We have two
possibilities depending on whether the inequality
$\frac{p_{m+2}}{q_{m+2}} \geq \frac{k}{d}$ is satisfied or not.

Assume first that  $\frac{p_{m+2}}{q_{m+2}} \geq \frac{k}{d}$. We
are searching for $\frac{k}{d}$ among the fractions of the form
$\frac{r'p_{m+3}+s'p_{m+2}}{r'q_{m+3}+s'q_{m+2}}$. As in Section
\ref{sec:VT}, we have $$ q_{m+2} >
\frac{\sqrt[4]{n}}{\sqrt{2.122(a_{m+3}+2)}}. $$ Now we have
\begin{eqnarray*}
 r' &=& dq_{m+2}\Big( \frac{p_{m+2}}{q_{m+2}} - \frac{k}{d} \Big) <
dq_{m+2} \cdot \frac{0.122 e}{n\sqrt{n}} < 0.061 dq_{m+2}\Big(
\frac{p_{m+2}}{q_{m+2}} - \frac{e}{n} \Big) \\ &<& 0.061
\frac{d}{q_{m+3}} < \frac{0.061 \sqrt{2.122(a_{m+3}+2)}}{a_{m+3}}
\,D
\end{eqnarray*}
and
\begin{eqnarray*} s' &=& dq_{m+3}\Big( \frac{k}{d} -
\frac{p_{m+3}}{q_{m+3}} \Big) \leq  dq_{m+3}\Big(
\frac{p_{m+2}}{q_{m+2}} - \frac{p_{m+3}}{q_{m+3}} \Big) =
\frac{d}{q_{m+2}} \\ &<& \sqrt{2.122(a_{m+3}+2)} \,D.
\end{eqnarray*}
Hence, $\frac{k}{d}$ can be recovered in at most $r's' <
\frac{0.1295(a_{m+3}+2)}{a_{m+3}} \,D^2 \leq 0.3885\,D^2$ steps.
Here $D=d/\sqrt[4]{n}$, as before.

\medskip

Assume now that $\frac{p_{m+2}}{q_{m+2}} < \frac{k}{d}$. We have
$$ \frac{k}{d} - \frac{e}{n} < \frac{2.122e}{n\sqrt{n}} <
\frac{2.122}{\sqrt{n}} = \frac{2.122D^2}{d^2}. $$ We are in the
conditions of the proof of Theorem \ref{tm:kbb}, and we conclude
that $\displaystyle{\frac{k}{d} =
\frac{rp_{m+1}+sp_{m}}{rq_{m+1}+sq_{m}}}$ or
$\displaystyle{\frac{k}{d} =
\frac{sp_{m+2}-tp_{m+1}}{sq_{m+2}-tq_{m+1}}}$, where $r$, $s$ and
$t$ are positive integers satisfying $rs < 4.244D^2$,
$st<4.244D^2$.

From the Dirichlet's formula for the number of divisors we obtain
immediately that the number of possible pairs $(r,s)$ and $(s,t)$
is $O(D^2\log{D})$. However, $r$ and $s$ (resp. $s$ and $t$) are
not arbitrary. They satisfy the inequalities $r<a_{m+2}s$ and
$t<a_{m+2}s$, which imply $r<2.061\sqrt{a_{m+2}}D$ and
$t<2.061\sqrt{a_{m+2}}D$. In Section \ref{sec:VT} we found that
$s\leq s_1$, where $s_1=\lfloor \sqrt{2.122(a_{m+2}+2)}D \rfloor$
if $\displaystyle{\frac{e}{n} - \frac{p_{m+1}}{q_{m+1}} >
\frac{2.122e}{n\sqrt{n}}}$, and $s_1=\lfloor
\sqrt{2.122(a_{m+2}+2)} \\ (a_{m+1}+1)D \rfloor$ if
$\displaystyle{\frac{e}{n} - \frac{p_{m+1}}{q_{m+1}} \leq
\frac{2.122e}{n\sqrt{n}}}$. Let $\displaystyle{s_0=\Big\lfloor
2.061 \frac{D}{\sqrt{a_{m+2}}} \Big\rfloor}$. We have the
following upper bound for the number of possible pairs $(r,s)$:
\begin{eqnarray*}
\lefteqn{a_{m+2}(1+2+\cdots+s_0) + \frac{D^2}{s_0+1} +
\frac{D^2}{s_0+2} + \cdots + \frac{D^2}{s_1}} \\ &\!<\!& a_{m+2}
s_0^2 + D^2(\log{\frac{s_1}{s_0+1}} +1) \\ &\!<\!& 5.248D^2 + D^2
\log(0.707 \max(\sqrt{(a_{m+3}\!+\!2)a_{m+2}},
(a_{m+2}\!+\!1)(a_{m+1}\!+\!1))).
\end{eqnarray*}
We have the same upper bound for the number of possible pairs
$(s,t)$.

Hence, the number of steps in this attack is $O(D^2\log{A})$
$(A=\max \{a_i \,:\, i=m+1,m+2,m+3\})$. We may compare this with
Verheul \& van Tilborg attack where the number of steps was
$O(D^2A^2)$.

\bigskip

\begin{example} \label{ex:2}
{\rm Let $n=7978886869909$, $e=4603830998027$, and assume that
$d<10000000$. Continued fraction expansion of $\frac{e}{n}$ is $$
[0, 1, 1, 2, 1, 2, 1, 18, 10, 1, 3, 3, 1, 6, 57, 2, 1, 2, 14, 7,
1, 2, 1, 4, 6, 2], $$ and the convergents are $$ 0,\, 1,\,
\frac{1}{2},\,  \frac{3}{5},\, \frac{4}{7},\, \frac{11}{19},\,
\frac{15}{26},\, \frac{281}{487},\, \frac{2825}{4896},\, \ldots
\,.$$ We find that $$\frac{281}{487} < \frac{e}{n} + \frac{2.122
e}{n\sqrt{n}} < \frac{11}{19}. $$ Hence $m=5$ and we are searching
for the secret exponent among the numbers of the form $26r+19s$ or
$487s - 26t$ or $4896r'+487s'$. By applying Wiener's test, we find
that $s=12195$, $t=77$ gives the correct value for $d$,
$d=5936963$.

Let us compare these numbers $s$ and $t$ with the numbers
$r$ and $s$ obtained by an application of the Verheul and van Tilborg
attack to the same problem. We obtain the same number $s=12195$, but the other number $r=219433$ is much larger than $t=77$,
which is in a good agreement with our theoretical estimates.

}
\end{example}

\bigskip

\begin{example} \label{ex:3}
{\rm Let us take $n=7978886869909$ again. For
$1000 \leq d \leq 1000000$, we compare the
quantities $rs$, obtained by Verheul and van Tilborg attack,
with the quantity $D^2$. The maximal value for $rs/D^2$ is $78464.2$
and it is attained for $d=611131$. There are 591
$d$'s for which $rs/D^2$ is greater than 1000.
The average value of
$rs/D^2$ for $d$ in the given interval is $15.69$.

Similar analysis for the attack introduced in this section gives
that the average value of the quantity $\min(rs,st,r's')/D^2$ for
$d$ in interval $1000 \leq d \leq 1000000$ is $0.8397$, with
maximal value 4.026 attained for $d=437561 $.

}
\end{example}

\bigskip

{\small \noindent Department of Mathematics \\ University of
Zagreb
\\ Bijeni\v cka cesta 30, 10000 Zagreb \\ Croatia \\
{\em E-mail address}: {\tt duje@math.hr}}

\end{document}